\documentclass[aps,prl,twocolumn,superscriptaddress,showpacs,floatfix,longbibliography]{revtex4-2}
\usepackage{amsmath,amssymb,amsfonts,float,graphics,epsfig,epstopdf,color,verbatim,tabularx,bm,multirow,appendix,hyperref}
\usepackage[normalem]{ulem}
\usepackage{lmodern}
\usepackage{ytableau}
\usepackage{pifont}
\usepackage{color}
\usepackage{bm}
\DeclareMathOperator{\Tr}{Tr}

\def\bk{{\mathbf{k}}}

\def\bK{{\mathbf{K}}}

\def\br{{\mathbf{r}}}

\def\bq{{\mathbf{q}}}

\def\bG{{\mathbf{G}}}

\begin{document}
	
\title{Quantum Monte Carlo sign bounds, topological Mott insulator and thermodynamic transitions in twisted bilayer graphene model}
\author{Xu Zhang}
\affiliation{Department of Physics and HKU-UCAS Joint Institute
	of Theoretical and Computational Physics, The University of Hong Kong,
	Pokfulam Road, Hong Kong SAR, China}

\author{Gaopei Pan}
\affiliation{Beijing National Laboratory for Condensed Matter Physics and Institute of Physics, Chinese Academy of Sciences, Beijing 100190, China
}
\affiliation{School of Physical Sciences, University of Chinese Academy of Sciences, Beijing 100049, China}

\author{Bin-Bin Chen}
\affiliation{Department of Physics and HKU-UCAS Joint Institute
	of Theoretical and Computational Physics, The University of Hong Kong,
	Pokfulam Road, Hong Kong SAR, China}

\author{Heqiu Li}
\affiliation{Department of Physics, University of Toronto, Toronto, Ontario M5S 1A7, Canada}
\author{Kai Sun}
\email{sunkai@umich.edu}
\affiliation{Department of Physics, University of Michigan, Ann Arbor, Michigan 48109, USA}

\author{Zi Yang Meng}
\email{zymeng@hku.hk}
\affiliation{Department of Physics and HKU-UCAS Joint Institute
	of Theoretical and Computational Physics, The University of Hong Kong,
	Pokfulam Road, Hong Kong SAR, China}

\begin{abstract}
We show that for magic-angle twisted bilayer graphene (TBG) away from charge neutrality, although quantum Monte Carlo (QMC) simulations suffer from the sign problem, the computational complexity is at most polynomial at certain integer fillings. For even integer fillings, this polynomial complexity survives even if an extra inter-valley attractive interaction is introduced, on top of Coulomb repulsions. This observation allows us to simulate magic-angle twisted bilayer graphene and to obtain accurate phase diagram and dynamical properties. At the chiral limit and filling $\nu=1$,
the simulations reveals a thermodynamic transition separating metallic state and a $C=1$ correlated Chern insulator -- topological Mott insulator (TMI) -- and the pseudogap spectrum slightly above the transition temperature. The ground state excitation spectra of the TMI exhibit a spin-valley U(4) Goldstone mode and a time reversal restoring excitonic gap smaller than the single particle gap. These results are qualitatively consistent with the recent experimental findings at zero-field and $\nu=1$ filling in $h$-BN nonaligned TBG devices.
\end{abstract}
\date{\today}
\maketitle

\noindent{\textcolor{blue}{\it Introduction.}---}
Magic-angle twisted bilayer graphene (TBG) has attracted great attention in recent years, as it hosts a variety of nontrivial phases beyond semi-classical or band-theory description~\cite{lopesGraphene2007,tramblyLocalization2010,rafiMoire2011,lopesContinuum2012,tramblyNumerical2012,rozhkovElectronic2016,caoUnconventional2018,caoCorrelated2018,xieSpectroscopic2019,luSuperconductors2019,kerelskyMaximized2019,liaoValence2019,yankowitzTuning2019,yankowitzTuning2019,tomarkenElectronic2019,caoStrange2020,kevinStrongly2020,soejimaEfficient2020,chatterjeeSkyrmion2022,khalafSoftmodes2020,xieNature2020,khalafCharged2021,pierceUnconventional2021,liaoCorrelated2021,rozenEntropic2021,zondinerCascade2020,saitoIsospin2021,parkFlavour2021,kwanExciton2021,liaoCorrelation2021,kangCascades2021,liuTheories2021,schindlerTrion2022,brillauxAnalytical2022,songMagic2022,linSpin2022,bhowmikBroken2022,huangObservation2022,zhangCorrelated2022,herzogReentrant2022,andreiGraphene2020,stepanovCompeting2021}. 
To theoretically characterize these flat bands and correlated quantum phases, tight-binding~\cite{tramblyLocalization2010,tramblyNumerical2012} and continuous (BM)~\cite{rafiMoire2011} models has been developed. In this study, we focus on the continuous-model approach, which avoids the challenge to construct localized orbitals that preserves all the symmetries~\cite{songMatbg2021,poOrigin2018,koshinoMaximally2018,kangSymmetry2018,poFragile2018,poFaithful2019}. By projecting long-range Coulomb interactions onto the moir\'e flat bands with proper quantum metric, such projected Hamiltonian has been studied using mean-field approximations~\cite{xieNature2020,liu2021nematic,liuTheories2021,zhangCorrelated2020}. At certain limits, exact analytical solution has also been obtained~\cite{BAB3,BAB4,BAB5,bultinckGround2020,vafekRenormalization2020}. On the numerical side, the charge neutrality point has been studied using sign-problem-free momentum-space quantum Monte Carlo (QMC) simulations~\cite{zhangMomentum2021,hofmannFermionic2022,panDynamical2021}. However, away from the charge neutrality point, due to the arising of the sign problem, such simulations have not yet been performed.


Although the sign problem often implies exponential computational complexity, it is worthwhile to emphasize that not all sign problems cause such severe damage.
Very recently, a much more mild type of sign problem has been demonstrated, where the computation complexity scales as a polynomial function of the system size, known as polynomial sign problem~\cite{ouyangProjection2021,panThermodynamic2022}.

In this Letter, we study the sign problem in TBG flat bands. Utilizing the sign bound theory~\cite{zhangFermion2022}, we prove that TBG flat bands at even integer fillings, or arbitrary integer fillings in the chiral limit, exhibit (at most) polynomial sign problem. This observation allows us to utilize QMC methods to study fillings away from charge neutrality, away from the chiral limit, and/or in the presence of extra attractive interactions on top of the Coulomb repulsion (See Tab.~\ref{tab1} for details).

To demonstrate this new approach, we performed large-scale QMC simulations to examine the chiral limit at filing $\nu=1$ (we denote the fully empty/filled flat bands as $\nu=-4/+4$ and charge neutrality as $\nu=0$). At $T=0$, 
this model can be solved exactly~\cite{BAB4,bultinckGround2020,hofmannFermionic2022}, and the exact solution reveals that at $T\to 0$, the system is a correlated Chern insulator -- a topological Mott insulator (TMI)~\cite{raghuTopological2008,chenRealization2021,linExciton2022,shenCorrelated2020,liuSpectroscopy2021,huangGiant2020} -- with Chern number $C=1$. Upon raising the temperature, the insulating state shall ``melt" into a metal and the time-reversal symmetry shall be recovered. However, our knowledge about this finite-temperature phase transition is very limited. Because the sign problem is only polynomial,  QMC simulations become a highly efficient tool to study this system, from which three different phases/states are observed (1) a metal phase with time-reversal symmetry at high temperature $T>T^\star$, (2)  a time-reversal invariant pseudo-gap phase at intermediate temperature $T_c<T<T^\star$ and (3) a low-temperature TMI phase at $T<T_c$. Here, $T^\star$ is a crossover temperature scale and $T_c$ is the critical temperature of a second order phase transition, at which the time-reversal symmetry is spontaneously broken. 
We further show that the TMI phase only breaks the time reversal symmetry, while spins remain disordered due to thermal   excitations of gapless spin fluctuations. This absence of spin order/polarization is in direct contrast to quantum Hall states where electron spins are polarized due to Zeeman splitting and thus spin fluctuations are gapped. 
In the discussion section, we establish the connection and consistency between our simulations and recent experimental studies on TMI phases in $h$-BN nonaligned TBG devices~\cite{stepanovCompeting2021}.

{\noindent \textcolor{blue}{\it BM model and projected interaction.}---}
We utilize the BM model and project interactions between fermions
to the moir\'e flat bands. The BM model Hamiltonian~\cite{rafiMoire2011} for the $\tau$ valley takes the following form
	$H^{\tau}_{\bk,\bk^\prime}
	= \left(\begin{array}{cc} 
		-\hbar v_F (\bk-\bK_1) \cdot \boldsymbol{\sigma} \delta_{\bk,\bk^\prime}  &  V  \\
		V^\dagger  & -\hbar v_F (\bk-\bK_2) \cdot \boldsymbol{\sigma}\delta_{\bk,\bk^\prime}
	\end{array}\right)$
where $\bK_1$ and $\bK_2$ mark the two Dirac points in the $\tau$ valley from layers $1$ and $2$ respectively.
$V=U_0\delta_{\bk,\bk^\prime}+U_1 \delta_{\bk,\bk^\prime-\bG_1}+U_2 \delta_{\bk,\bk^\prime-\bG_1-\bG_2}$ and 
$	V^\dagger=U_0^\dagger\delta_{\bk,\bk^\prime}+U_1^{\dagger} \delta_{\bk,\bk^\prime+\bG_1}+U_2^{\dagger} \delta_{\bk,\bk^\prime+\bG_1+\bG_2}$ are the inter-layer tunnelings with matrixes $U_0=
\left(\begin{array}{cc} 
	u_0  & u_1 \\ 
	u_1 & u_0
\end{array}\right)$, $U_1=\left(\begin{array}{cc} 
	u_0  & u_1 e^{-i\frac{2\pi}{3}} \\ 
	u_1 e^{i\frac{2\pi}{3}} & u_0
\end{array}\right)$ and $U_2=\left(\begin{array}{cc} 
	u_0  & u_1 e^{i\frac{2\pi}{3}} \\ 
	u_1 e^{-i\frac{2\pi}{3}} & u_0
\end{array}\right)$
where $u_0$ and $u_1$ are the intra- and inter-sublattice inter-layer tunneling amplitudes. 
$\bG_1=(-1/2,-\sqrt{3}/2), \bG_2=(1,0)$ are the reciprocal vectors of the moir\'e Brillouin zone (mBZ), and 
$\bK_1=(0,1/2\sqrt{3})|\bG_{1,2}|, \bK_2=(0,-1/2\sqrt{3})|\bG_{1,2}|$.
For control parameters, we set $(\theta,\hbar v_F/a_0,u_0,u_1)=(1.08^{\circ},2.37745\,\mathrm{eV},0\,\mathrm{eV},0.11\,\mathrm{eV})$ for the chiral limit, and for non-chiral model, we set $u_0=0.06$ eV following Refs.~\cite{BAB1,BAB3,zhangMomentum2021,panDynamical2021}. Here, $\theta$ is the twisting angle and $a_0$ is the lattice constant of monolayer graphene.



For interactions, in addition to the Coulomb repulsion, here we have the option to include one more interaction term and the sign problem will still remain polynomial.
\begin{align}
	H_{I}=&\frac{1}{2 \Omega} \sum_{\mathbf{q}} \left( V_1(\mathbf{q}) \delta \rho_{1,\mathbf{q}} \delta \rho_{1,\mathbf{-q}} + V_2(\mathbf{q}) \delta \rho_{2,\mathbf{q}} \delta \rho_{2,\mathbf{-q}}\right) \nonumber\\
	\delta \rho_{1,\mathbf{q}}=&\sum_{\mathbf{k},\alpha,\tau,s} (c_{\mathbf{k},\alpha,\tau,s}^{\dagger} c_{\mathbf{k+q},\alpha,\tau,s}-\frac{\nu+4}{8}\delta_{\mathbf{q},0}) \nonumber\\
	\delta \rho_{2,\mathbf{q}}=&\sum_{\mathbf{k},\alpha,s} (c_{\mathbf{k},\alpha,\tau,s}^{\dagger} c_{\mathbf{k+q},\alpha,\tau,s}-c_{\mathbf{k},\alpha,-\tau,s}^{\dagger} c_{\mathbf{k+q},\alpha,-\tau,s})
	\label{eq:eq2}
\end{align}
The first term in $H_I$ ($V_1>0$) is the Coulomb interactions, and the second term $V_2\ge 0$ introduces repulsive interactions for fermions in the same valley and attractions between the two valleys, which can be introduced as a phenomenology term describing inter-valley attractions~\cite{Ge2013phonon,Rold2013Interactions,Yuan2014Possible,zhang2021superconductivity,anInteraction2020}.
At $V_2=0$, this model recovers the standard TBG model with Coulomb repulsions. When $V_2$ is turned on, the inter-valley attraction favors inter-valley pairing and could stabilize a superconducting ground state. 
The normalization factor in $H_I$ is $\Omega=L^2 \frac{\sqrt{3}}{2} a_{M}^{2}$ with $L$ being the linear system size of the system. $\mathbf{k}$ and $\mathbf{q}$ cover the whole momentum space, $\nu$ is the filling factor and $\alpha,\tau,s$ represent layer/sublattice, valley, spin indices, respectively. 
The mometum dependence for non-negative $V_1$ and $V_2$ is unimportant to polynomial sign problem.
Here, for simplicity, we set $V_2=\gamma V_1$ with $\gamma$ being a non-negative constant and for $V_1$, we use Coulomb interaction screened by single gate in our simulation $V_1(\bq) =\frac{e^{2}}{4 \pi \varepsilon} \int d^{2} \br\left(\frac{1}{\br}-\frac{1}{\sqrt{\br^{2}+d^{2}}}\right) e^{i \bq \cdot \br}=\frac{e^{2}}{2 \varepsilon} \frac{1}{\left| \bq\right| }\left(1-e^{-\left| \bq\right|  d}\right)$ where $\frac{d}{2}=20\,\mathrm{nm}$ is the distance between graphene layer and single gate, $\epsilon=7\epsilon_0$ is the dielectric constant. We then project the interactions $H_I$ to the moir\'e flat bands (See SM~\cite{suppl}) and use the projected Hamiltonian to carry out sign bounds analysis and QMC simulations.

{\noindent \textcolor{blue}{\it Polynomial sign bounds.}---}
In QMC simulations, the expectation value of a physical observable $O$ is measured as
$\langle \hat O \rangle =\sum_l W_l \langle \hat O\rangle_l$, where $W_l$ and $\langle \hat O\rangle_l$ are the weight and the expectation value for the configuration $l$. 
Instead of summing over all configurations, a QMC simulation samples the configuration space using the probability $W_l$.  In sign-problem-free QMC simulations, $W_l \ge 0$ for all $l$ and an accurate expectation value can be obtained by only sampling a small number of configurations -- the importance sampling, and this number scales as a power-law function of the system size. However, for many quantum systems, $W_l$ can be negative or even complex, and thus to obtain an accurate expectation value, it requires to sample a large number of configurations, which usually scales as an exponential function of the system size~\cite{panSign2022}.

It is worthwhile to emphasize that the sign problem doesn't always lead to an exponentially high computational cost. To measure the severity of the sign problem, here we use the average sign
$\left\langle sign \right\rangle = \sum_l W_l / \sum_l |Re(W_l)|$, where $|Re( W_l)|$ is the absolute value of the real part of $W_l$.
For physical partition function $Z=\sum_l W_l=\sum_l Re(W_l)$, this average sign is between $0$ and $1$, and $\left\langle sign \right\rangle=1$ means that the system is sign problem free, while smaller $\left\langle sign \right\rangle$ means severer sign problem. In a $d$-dimension quantum systems that suffers from the sign problem, $\left\langle sign \right\rangle \sim \exp(-\beta L^d)$ where $\beta=1/T$ the inverse temperature,  indicating that the number of configurations needed in QMC simulations scales as an exponential function of the space-time volume. For polynomial sign problem, although $\left\langle sign \right\rangle<1$ (i.e., the system does suffer from the sign problem),  $1/\left\langle sign \right\rangle$ is  a polynomial  function of the system size, and thus the number of configurations needed only scales as a power-law function of the system size. 

Although the average sign  can be easily measured in QMC simulations, it usually doesn't have a simple analytic formula. 
To estimate the numerical cost to overcome the sign problem, we utilize the sign bound $\left\langle sign \right\rangle_b$ defined in Ref.~\cite{zhangFermion2022}. As proved in Ref.~\cite{zhangFermion2022}, $\left\langle sign \right\rangle_b$ is the lower bound of $\left\langle sign \right\rangle$ (i.e., $\left\langle sign \right\rangle_b\leqslant\left\langle sign \right\rangle$). Thus, if the sign bound scale is a power-law function of the system size, the sign problem is (at most) polynomial.
Remarkably, the low temperature sign bound in moir\'e flat bands can be easily calculated by counting ground state degeneracy, which can be obtained using SU(4) and SU(2) Young diagram as employed in Refs.~\cite{BAB3,BAB4,hofmannFermionic2022} (See SM for details~\cite{suppl}). 

Details about this calculation are shown in the SM and  the conclusions are summarized in Tab.~\ref{tab1}. 
At charge neutrality ($\nu=0$), moir\'e flat bands with Coulomb interactions ($\gamma=0$) is known to be sign problem free~\cite{zhangMomentum2021,hofmannFermionic2022}, and thus the sign bound is $1$. Here, we further prove that adding inter-valley attractions ($\gamma >0$) to the chiral system doesn't cause sign problem either ($\left\langle sign \right\rangle_b=1$). Away from charge neutrality,  sign problem arises, but it is polynomial at certain integer fillings. For  Coulomb repulsion ($\gamma=0$), the sign problem is polynomial at any (even) integer fillings at (away from) the chiral limit. When inter-valley attractions are introduced ($\gamma>0$), even integer fillings at the chiral limit also have polynomial sign bound.

\begin{table}[t]
	\caption{Scaling of the sign bound $\left\langle sign\right\rangle_b $ at low temperature and large moir\'e lattice size $N=L^2$. A power-law function of $N$ indicates that the sign problem is (at most) polynomial. \ding{55} indicates the sign bound decays to zero exponentially.}
	\begin{ruledtabular}
		\begin{tabular}{cccc}
			\textrm{Filling($\nu$)}&
			\textrm{Chiral($\gamma=0$)}&
			\textrm{Non-chiral($\gamma=0$)}&
			\textrm{Chiral($\gamma>0$)}\\
			\colrule
			0 & 1 & 1 & 1 \\
			$\pm1$ & $N^{-1}$ & \ding{55} & \ding{55}\\
			$\pm2$ & $N^{-2}$ & $N^{-1}$ & $N^{-2}$\\
			$\pm3$ & $N^{-5}$ & \ding{55} & \ding{55}\\
			$\pm4$ & $N^{-8}$ & $N^{-4}$ & $N^{-4}$\\
		\end{tabular}
	\end{ruledtabular}
	\label{tab1}
\end{table}

\begin{figure}[htp!]
	\includegraphics[width=1.0\columnwidth]{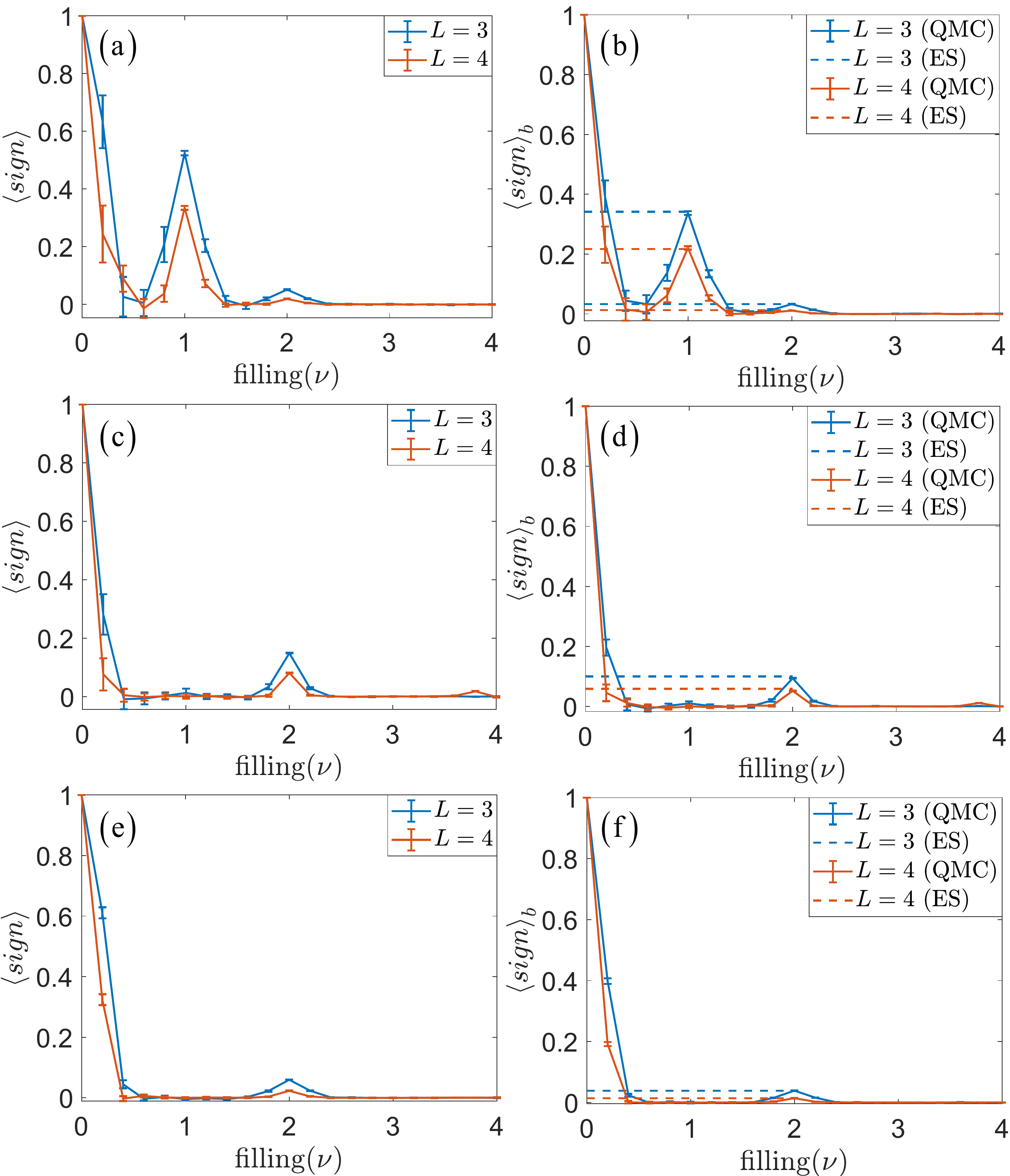}
	\caption{$\langle sign \rangle$ versus filling $\nu$ (a,c,e) and $\langle sign \rangle_b$ versus filling $\nu$ (b,d,f) at low temperature $T=1$ meV. The fillings of $\nu<0$ are symmetric with respect to $\nu=0$ via the particle-hole symmetry. (a-b) are the chiral limit $\gamma=0$ cases, (c-d) are the non-chiral $\gamma=0$ cases where we take $u_0=0.06\,\mathrm{eV}$ and (e-f) are the chiral limit $\gamma=4$ cases. (a,c,d) are the average sign $\left\langle sign \right\rangle $ for $L=3 (N=9)$ and $L=4 (N=16)$ measured from QMC for filling from $\nu=0$ to $\nu=4$. (b,d,f) are the sign bounds $\left\langle sign \right\rangle_b $ for $L=3 (N=9)$ and $L=4 (N=16)$ measured from QMC (solid line) and derived from exact solution (ES) at the low temperature limit for filling $\nu=1,2$ (dash line values). The ES values in (b) are 1119/3278, 3630/111493 for $L=3$ at $\nu=1,2$ and 945/4357, 183/14912 for $L=4$ at $\nu=1,2$. The ES values are 1/10 for $L=3$, 1/17 for $L=4$ at $\nu=2$ in (d) and 100/2601 for $L=3$, 71/5201 for $L=4$ at $\nu=2$ in (f). (See SM for details~\cite{suppl})}
	\label{fig:fig1}
\end{figure}

To further verify the polynomial sign problem summarized in Tab.~\ref{tab1}, we directly calculate $\langle sign \rangle$ and $\langle sign \rangle_b$ in QMC simulations at various filling $\nu$, and compare them with the exact formula of the sign bound obtained at integer fillings. As shown in Fig.~\ref{fig:fig1},  $\langle sign \rangle$  is always larger or equal to
$\langle sign \rangle_b$ as expected, and the sign bound at integer filling indeed converges to the exact solution. 
The peak in $\langle sign \rangle$ at integer (or even integer) fillings indicates that the sign problem is less severe and QMC simulations have a faster convergence at these fillings. The location of these peaks are fully consistent with the polynomial sign problem summarized in Tab.~\ref{tab1}. The only exceptions are $\nu=3$ and $4$ of Fig.~\ref{fig:fig1}(a), where the sign problem is polynomial but the figure doesn't exhibit visible peaks. This is because the power-law function here has high powers $N^{-5}$ and $N^{-8}$, which requires higher resolution (longer simulation time) to show clear distinction from exponential functions. 


{\noindent \textcolor{blue}{\it Chiral limit at $\nu=1$ and $\gamma=0$.}---}
With the polynomial sign bound obtained, here we discuss the QMC results as a function of temperature for the chiral limit $\nu=1$ filling case in this section. In the QMC simulation, we observe a thermal phase transition with pseudogap spectrum and spontaneous time reversal symmetry breaking, which is of immediate relevance to the recent experimental finding of correlated Chern insulators at zero-field and $\nu=1$ filling in $h$-BN nonaligned TBG devices and its relatively high Curie temperature of $T_c \sim 4.5 $ K~\cite{stepanovCompeting2021}.

At low temperature limit for $\nu=1$, exact solution at $T=0$ expects degenerate ground states with Chern number $C=\pm1$, and $\pm3$~\cite{BAB4,bultinckGround2020,hofmannFermionic2022}. In the large system size limit $N\to \infty$, the number of ground states scales as $\propto N^7$ for $C=\pm 1$ and $N^3$ for $C=\pm 3$~\cite{suppl}. Due to the higher number of ground state degeneracy, thermal fluctuations shall stabilize the $C=\pm1$ state as the thermal equilibrium state at low temperature via the order by disorder mechanism~\cite{henleyOrdering1989}. This low-temperature state breaks spontaneously the time-reversal symmetry,  but this symmetry breaking process as a function of temperature is unknown.

\begin{figure}[t]
	\includegraphics[width=1.0\columnwidth]{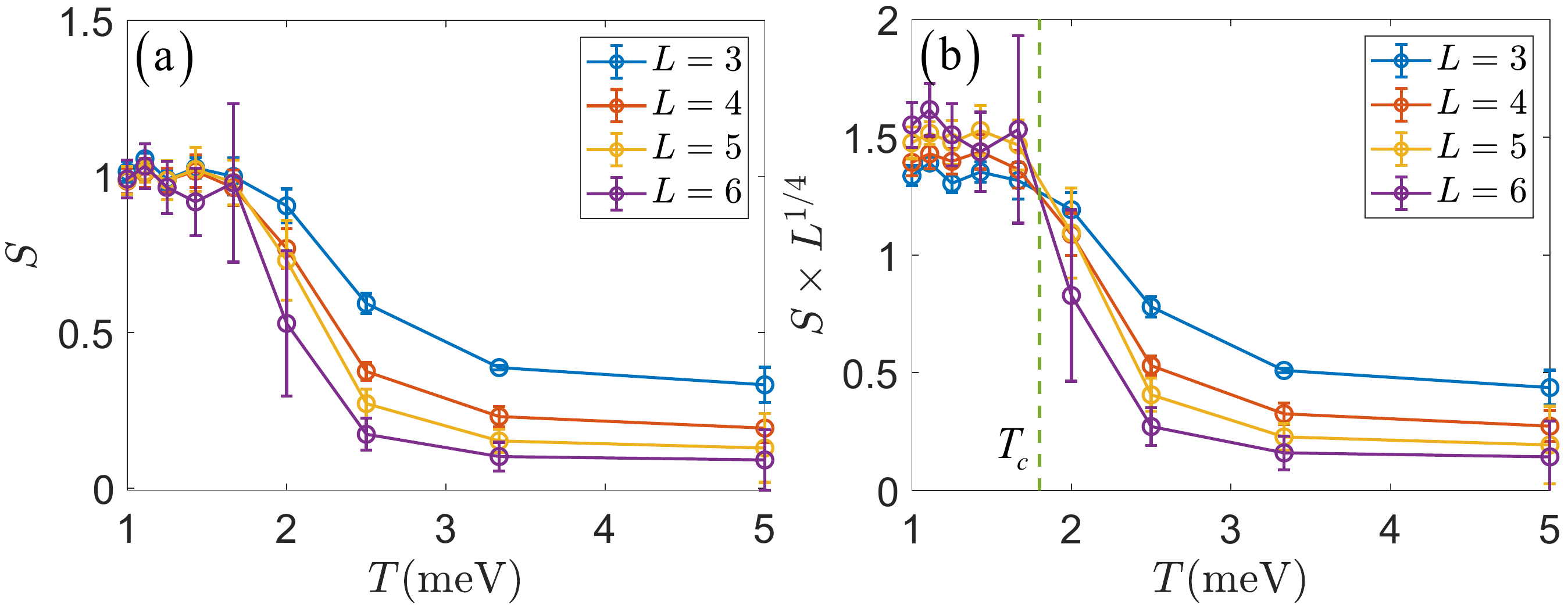}
	\caption{(a) Chern band polarization correlation function $S$ versus $T$. At low temperature, the Chern number approaches to 1. (b) 2D Ising universality class crossing to determine the phase transition point at $T_c\sim1.8$ meV, denoted by the dashed line.}
	\label{fig:fig2}
\end{figure}

Our QMC simulation at finite temperature reveals this process. To probe the time-reversal symmetry breaking, we use the Chern band polarization as the order parameter, $\langle \hat{N}_+ - \hat{N}_-\rangle/N$, where $\hat{N}_\pm$ are the fermion occupation number operators of $\pm$ Chern bands. The correlation function of this order parameter is plotted in Fig.~\ref{fig:fig2}, $S\equiv\langle (\hat{N}_+ - \hat{N}_-)^2 \rangle/N^2$, and scaling analysis reveals a second order phase transition at $T_c\sim1.8$ meV, below which the time-reversal symmetry is spontaneously broken, similar to what was observed at $\nu=3$ with real-space effective models~\cite{chenRealization2021,linExciton2022,panThermodynamic2022}.  
At low temperature, $S$ approaches $1$, indicating that the Chern number is $C= \pm 1$, instead of $\pm 3$. In Fig.~\ref{fig:fig2}(b), we rescale the $y$ axis as $S\times L^{2\beta/\nu}$ using the 2D Ising exponents $\beta=1/8$ and $\nu=1$, and the cross point at $T_c\sim1.8$ meV marks the critical temperature.

\begin{figure}[t]
	\includegraphics[width=1.0\columnwidth]{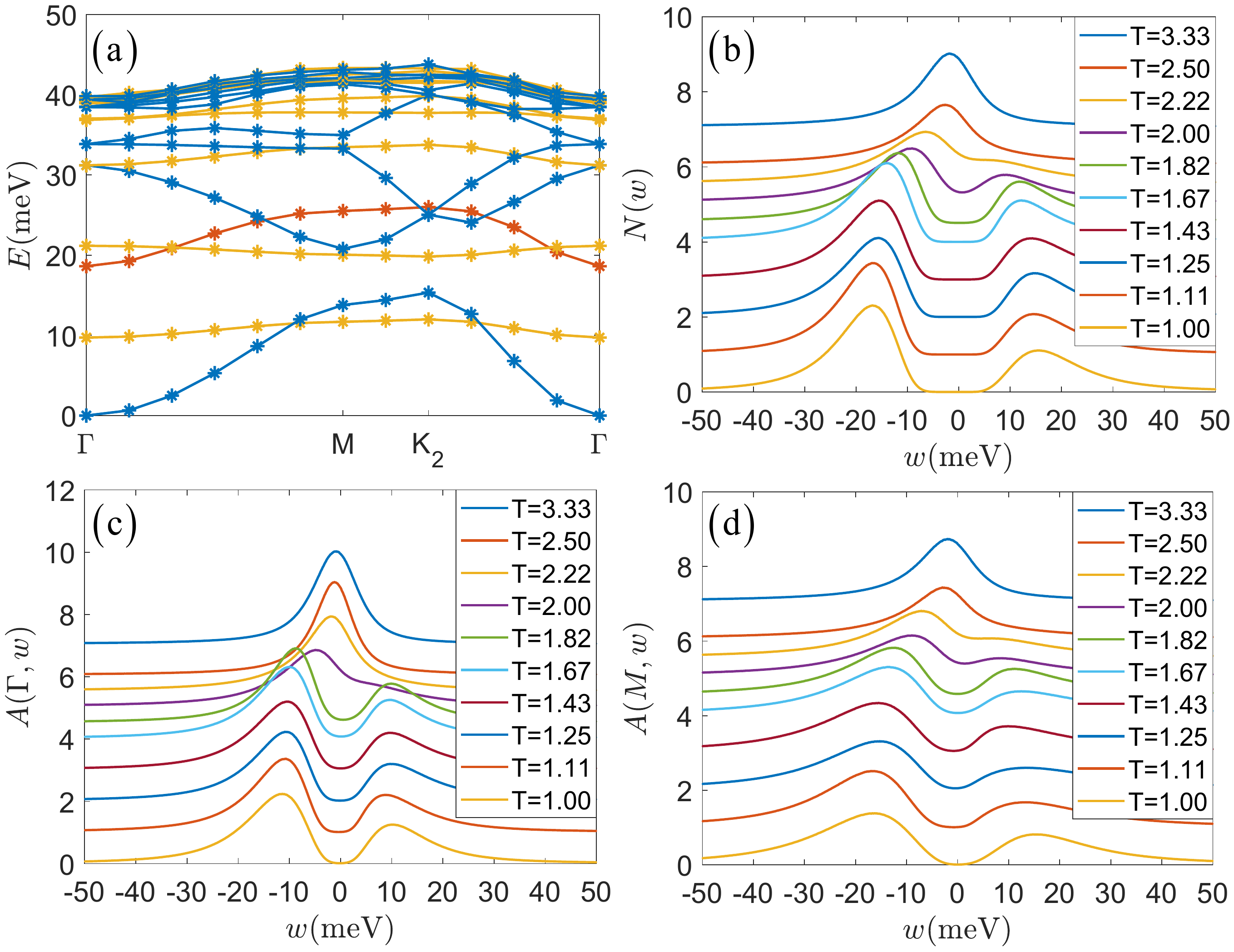}
	\caption{(a) Analytical excitation spectra along a high symmetry path in mBZ. The red line labels the single particle excitations. Blue lines show the 10 lowest charge neutral excitons between bands with the same Chern number. 
	The yellow lines label the 10 lowest charge neutral excitons between opposite Chern bands, which are responsible to the restoration of time reversal symmetry.
(b-d) Local density of states $N(\omega)$, the single particle spectra $A(\Gamma,\omega)$ and $A(M,\omega)$ obtained from the QMC-SAC scheme as a function of $T$. Different lines are lifted in $y$ direction with the amount of $\Delta\beta=\Delta(1/T)$ for clarity. Between the high temperature metal-like phase and the low temperature TMI phase, pseudo gap behavior can be find at $T^\star \sim 2$ meV above $T_c=1.8$ meV.}
	\label{fig:fig3}
\end{figure}

Such a spontaneously generated Chern insulator is of both theoretical and experimental interesting as in the temperature range of $0<T<T_c$, it only breaks the time reversal symmetry but not the spin-valley U(4) continuous symmetry. In contrast to quantum Hall states, where fermion spins are polarized due to Zeeman splitting, spin degrees of freedom doesn't form any order in this TMI phase and the spin SU(2) symmetry is preserved. To better demonstrate this point, we follow the method in Ref.~\cite{BAB5} and analytically compute the spectrum of single-particle and charge-neutral excitonic excitations at $T=0$. As shown in Fig.~\ref{fig:fig3}(a), single particle excitations (red stars) are fully gaped, indicating an insulating state. To restore the time-reversal symmetry, it requires to move fermions from + Chern bands to - Chern bands (or vice versa). However, such particle-hole excitations are fully gapped (yellow stars in Fig.~\ref{fig:fig3}(a)), and thus thermal fluctuations at low $T$ cannot restore the time-reversal symmetry. In contrast, particle-hole excitations between spin-valley bands with the same Chern number (blue starts) are gapless. These excitons describe spin SU(2) and spin-valley SU(4) fluctuations, and any spin or spin-valley order would be destroyed by these gapless excitations at any finite temperature. 

We note that the energy scale of single particle gap is larger than the gap of time-reversal-restoring excitons [Fig.~\ref{fig:fig3}(a)], indicating that time-reversal symmetry breaking is probably more vulnerable to thermal fluctuations in comparison to single particle excitations. 
To probe this physics, we employ the stochastic analytic continuation (SAC) method upon the QMC imaginary time data of the Green's function to extract the real-frequency single-particle spectra~\cite{sandvikStochastic1998,beachIdentifying2004,sandvikConstrained2016,syljuaasenUsing2008}. 
Such QMC+SAC scheme has been successfully employed in many quantum many-body systems~\cite{sunDynamical2018,maDynamical2018,wangFractionalized2021,jiangMonte2022,zhouEvolution2022} including TBG and other moir\'e systems~\cite{zhangMomentum2021,panDynamical2021,zhang2021superconductivity} and compared well with exact solutions and various spectroscopic experiments~\cite{sandvikConstrained2016,shaoNearly2017,zhouAmplitude2021,liKosterlitz2020,huEvidence2020}.

The obtained local density of states (LDOS) $N(\omega)$ and the single particle spectral functions $A(\Gamma,\omega)$ and $A(M,\omega)$ are shown in Fig.~\ref{fig:fig3}(b-d), respectively. From them, one sees that slightly above the $T_c=1.8$ meV, at $T^\star=2$ meV, the spectra indeed develop a pseudogap shape at both momenta and $N(\omega)$. Below $T_c$, the spectra are fully gapped and the system is an interaction-driven topological Mott insulator~\cite{raghuTopological2008,chenRealization2021,linExciton2022} with no spin polarization and Chern number $C=1$. The pseudogap behavior at $T_c<T<T^\star$, is certainly beyond mean-field description of the system, which would require the gap opens exactly at the transition, and is the manifestation of the intricate competition between the single-particle, collective excitations (such as the excitons) and thermal fluctuations in the morie system. Our LDOS results at low temperature with the asymmetric spectral weights, are also consistent with STM experiment at $\nu=1$~\cite{kerelskyMaximized2019}.


{\noindent \textcolor{blue}{\it Discussion.}---} The experimental observation of the zero-field Chern insulators in TBG, at $\nu=1$ and with $C=1$~\cite{stepanovCompeting2021}, clearly posts the question of how to understand the rich physics in pristine TBG systems, but it is known that the model level computations taking into account of the strong interaction and topological ingredient of the flat band wavefunctions at finite temperature, are notoriously challenging. Here we find the way out by using the fermion sign bounds theory~\cite{zhangFermion2022}, upon which we prove for Coulomb interaction and chemical potential projected on flat bands TBG model, all integer fillings at chiral and even integer fillings at non-chiral cases have either no sign problem or polynomial sign bounds in their QMC simulations. The similar behavior also retains when projected effective attraction is introduced for chiral even integer fillings. 

This new approach allows us to unbiasedly compute the physical properties of the model at finite temperature. 
For $\nu=1$, the numerical results are fully consistent with experimental observations, including spontaneous time-reversal symmetry breaking, Chern number $C=1$, and the asymmetric in LDOS.
For $T_c<T<T^\star$, the simulation reveals a pseudo gap phase. This result is consistent with the observation of insulating-like behavior at $T>T_c$ ~\cite{kerelskyMaximized2019}. This pseudo gap phase would be an interesting subject for further experimental studies, which will help us better understand the phase transition and mechanism that drives the time-reversal symmetry breaking in moir\'e flat bands. 

\begin{acknowledgements}
{\noindent \it Acknowledgements.---}We thank Wang Yao for discussions on the related topic. XZ, GPP, BBC and ZYM acknowledge support from the Research Grants
Council of Hong Kong SAR of China (Grant Nos.~17303019, 17301420, 17301721, AoE/P-701/20 and 17309822), the K. C. Wong Education Foundation (Grant No.~GJTD-2020-01), and the Seed Funding ``Quantum-Inspired explainable-AI'' at the HKU-TCL Joint Research Centre for Artificial Intelligence. We thank HPC2021 system under the Information Technology Services and the Blackbody HPC system at the Department of Physics, the University of Hong Kong for their technical support and generous allocation of CPU time.
\end{acknowledgements}

\bibliographystyle{apsrev4-2}
\bibliography{Chiral_TBG}

\newpage
\begin{widetext}	
\section{Supplemental Material for  \\[0.5em] Quantum Monte Carlo sign bounds, topological Mott insulator and thermodynamic transitions in twisted bilayer graphene model }

\section{QMC simulation}

We introduce momentum space QMC method and fermion sign bounds theory briefly here following~\cite{zhangMomentum2021,zhangFermion2022}. Starting from flat band Hamiltonian
\begin{eqnarray}
	H_{I}&=&\frac{1}{2 \Omega} \sum_{\mathbf{G}} \sum_{\mathbf{q} \in mBZ} \left( V_1(\mathbf{q+G}) \delta \rho_{1,\mathbf{q+G}} \delta \rho_{1,\mathbf{-q-G}} + V_2(\mathbf{q+G}) \delta \rho_{2,\mathbf{q+G}} \delta \rho_{2,\mathbf{-q-G}}\right) \nonumber\\
	\delta \rho_{1,\mathbf{q+G}}&=&\sum_{\mathbf{k},m,n,\tau,s} \lambda_{m,n,\tau}(\mathbf{k}, \mathbf{k+q+G}) (c_{\mathbf{k}, m, \tau,s}^{\dagger} c_{\mathbf{k+q}, n, \tau,s}-\frac{\nu+4}{8}\delta_{\mathbf{q},0}\delta_{m,n})\nonumber\\
	\delta \rho_{2,\mathbf{q+G}}&=&\sum_{\mathbf{k},m,n,s} \lambda_{m,n,\tau}(\mathbf{k}, \mathbf{k+q+G}) c_{\mathbf{k}, m, \tau,s}^{\dagger} c_{\mathbf{k+q}, n, \tau,s} - \lambda_{m,n,-\tau}(\mathbf{k}, \mathbf{k+q+G})c_{\mathbf{k}, m, -\tau,s}^{\dagger} c_{\mathbf{k+q}, n, -\tau,s}
\end{eqnarray}
Here $m,n$ are the band indexes and $\lambda_{m,n,\tau}(\mathbf{k}, \mathbf{k+q+G})=\sum_{\mathbf{G}^{\prime}, \alpha} u_{m, \tau ; \mathbf{G}^{\prime}, \alpha}^{*}(\mathbf{k}) u_{n, \tau ; \mathbf{G}^{\prime}+\mathbf{G}, \alpha}(\mathbf{k+q})$ is the so called form factor which is defined as the overlap of two eigen functions of bands $m,n$ at momenta $\bk,\bk+\bq$ and valley $\tau$. Discarding the flat band dispersion (take the flat band limit), truncating the summation over band index and only keeping the two flat bands, we arrive at the final model Hamiltonian above to carry out sign bounds analysis and QMC simulations. 

According to the discrete Hubbard-Stratonovich transformation, $e^{\epsilon \hat{O}^{2}}=\frac{1}{4} \sum_{l=\pm 1,\pm 2} \gamma(l) e^{\sqrt{\epsilon} \eta(l) \hat{o}}+O\left(\epsilon^{4}\right)$, 
where $\epsilon$ is a small constant, $\gamma(\pm 1)=1+\frac{\sqrt{6}}{3}$, $\gamma(\pm 2)=1-\frac{\sqrt{6}}{3}$, $\eta(\pm 1)=\pm \sqrt{2(3-\sqrt{6})}$ and $\eta(\pm 2)=\pm \sqrt{2(3+\sqrt{6})}$, we can rewrite the partition function as,

\begin{eqnarray}
	&Z&=\Tr\{\prod_{t}e^{-\Delta \tau H_{I}(t)}\} =\Tr\{\prod_{t} e^{-\Delta \tau \frac{1}{4 \Omega}\sum_{|\bq+\bG|\neq0,\alpha=1,2} V_\alpha(\bq+\bG)\left[\left(\delta\rho_{\alpha,-\bq-\bG}+\delta\rho_{\alpha,\bq+\bG}\right)^{2}-\left(\delta\rho_{\alpha,-\bq-\bG}-\delta\rho_{\alpha,\bq+\bG}\right)^{2}\right]  }\} \nonumber\\
	&\approx& \sum_{\{l_{\alpha,|\bq|,t}\}} \prod_{t} [ \prod_{|\bq|\neq0,\alpha}\frac{1}{16} \gamma\left(l_{\alpha,|\bq|_1,t}\right) \gamma\left(l_{\alpha,|\bq|_2,t}\right)]  \Tr\{\prod_{t}[\prod_{|\bq|\neq0,\alpha}e^{i \eta\left(l_{\alpha,|\bq|_1,t}\right) A_{\alpha,\bq}\left(\delta\rho_{\alpha,-\bq}+\delta\rho_{\alpha,\bq}\right)} e^{\eta\left(l_{\alpha,|\bq|_2,t}\right) A_{\alpha,\bq}\left(\delta\rho_{\alpha,-\bq}-\delta\rho_{\alpha,\bq}\right)}]\} \nonumber\\
\end{eqnarray}
In the last line we replace $\bq+\bG$ by $\bq$ for simple. Here $t$ is the imaginary time index with step $\Delta\tau$, $A_{\alpha,\bq} \equiv\sqrt{\frac{\Delta \tau}{4} \frac{V_\alpha(\bq)}{\Omega}}$ and $\{l_{1,|\bq|_1,t},l_{1,|\bq|_2,t},l_{2,|\bq|_1,t},l_{2,|\bq|_2,t}\}$ are $2N_q$ four-component auxiliary fields where $N_q$ is the amount of momentum points we consider. According to Ref.~\cite{BAB1,zhangMomentum2021}, we only keep momentum points within length of reciprocal vector $G$.

Generally, average of any observable $\hat{O}$ can be written as,

\begin{equation}
	\langle \hat{O} \rangle = \frac{\Tr(\hat{O} e^{-\beta H})}{\Tr(e^{-\beta H})} = \sum_{\{l_{\alpha,|\bq|,t}\}} \frac{P(\{l_{\alpha,|\bq|,t}\}) \Tr[\prod_{t}\hat{B}_t(\{l_{\alpha,|\bq|,t}\})] \frac{\Tr[\hat{O} \prod_{t}\hat{B}_t(\{l_{\alpha,|\bq|,t}\})]}{\Tr[\prod_{t}\hat{B}_t(\{l_{\alpha,|\bq|,t}\})]}}{\sum_{\{l_{\alpha,|\bq|,t}\}} P(\{l_{\alpha,|\bq|,t}\}) \Tr[\prod_{t}\hat{B}_t(\{l_{\alpha,|\bq|,t}\})]} 
\end{equation}
Here $P(\{l_{\alpha,|\bq|,t}\})\equiv\prod_{t} [ \prod_{|\bq|\neq0,\alpha}\frac{1}{16} \gamma\left(l_{\alpha,|\bq|_1,t}\right) \gamma\left(l_{\alpha,|\bq|_2,t}\right)]$ and $\hat{B}_t(\{l_{\alpha,|\bq|,t}\})\equiv\prod_{|\bq|\neq0,\alpha}e^{i \eta\left(l_{\alpha,|\bq|_1,t}\right) A_{\alpha,\bq}\left(\delta\rho_{\alpha,-\bq}+\delta\rho_{\alpha,\bq}\right)} e^{\eta\left(l_{\alpha,|\bq|_2,t}\right) A_{\alpha,\bq}\left(\delta\rho_{\alpha,-\bq}-\delta\rho_{\alpha,\bq}\right)}$, respectively. We see $W_l\equiv P(\{l_{\alpha,|\bq|,t}\}) \Tr[\prod_{t}\hat{B}_t(\{l_{\alpha,|\bq|,t}\})]$ as possibility weight and $\langle \hat{O} \rangle_l=\frac{\Tr[\hat{O} \prod_{t}\hat{B}_t(\{l_{\alpha,|\bq|,t}\})]}{\Tr[\prod_{t}\hat{B}_t(\{l_{\alpha,|\bq|,t}\})]}$ as sample value for configuration $\{l_{\alpha,|\bq|,t}\}$. Then Markov chain Mento Carlo can compute this $\langle \hat{O} \rangle$.

But here $W_l$ is not non-negative for all configurations, we need do reweighting to derive a non-negative weight as a well-defined possibility.
\begin{equation}
	\langle \hat{O} \rangle = \frac{\sum_{l} W_l\langle \hat{O} \rangle_l}{\sum_{l} W_l} = \frac{\frac{\sum_{l} |Re(W_l)|\frac{W_l\langle \hat{O} \rangle_l}{|Re(W_l)|}}{\sum_{l} |Re(W_l)|}}{\frac{\sum_{l} W_l}{\sum_{l} |Re(W_l)|}} \equiv \frac{\langle \hat{O} \rangle_{|Re(W_l)|}}{\left\langle sign \right\rangle }
\end{equation}
The final numerator is the measurement with reweighting possibility distribution and the denominator $\left\langle sign \right\rangle\equiv \frac{\sum_{l} W_l}{\sum_{l} |Re(W_l)|}$ is the well-known average sign. By noticing $\sum_{l} |Re(W_l)|\leqslant\sum_{l} |W_l|=Z_{\nu=0}$, we can define the lower bounds of sign as $\left\langle sign \right\rangle_b\equiv \frac{\sum_{l} W_l}{\sum_{l} |W_l|}$, which is also a well-defined observable expressed by partition functions of two systems $\frac{\sum_{l} W_l}{\sum_{l} |W_l|}=\frac{Z_{\nu}}{Z_{\nu=0}}$. \textit{We can discuss properties of partition functions by measuring sign bounds, and we also can forecast average sign behavior by analyzing partition functions.} This is the brief spirit of fermion sign bounds theory~\cite{zhangFermion2022}.

One useful result can be derived from the observation of all $\hat{B}_t(\{l_{\alpha,|\bq|,t}\})$ operators are unitary, so that for Green's function at valley $\tau$ defined as
\begin{equation}
	G_{i,j}(\tau)=\frac{\operatorname{Tr}\left[c_{i,\tau} c_{j,\tau}^\dagger \prod_{t} \hat{B}_{t,\tau}\left(\left\{l_{\alpha,|\bq|, t}\right\}\right) \right]}{\operatorname{Tr}\left[\prod_{t} \hat{B}_{t,\tau}\left(\left\{l_{\alpha,|\bq|, t}\right\}\right)\right]}=[(I+U)^{-1}]_{i,j}
\end{equation}
we have $G(\tau)+G^{\dagger}(\tau)=(I+U)^{-1}+(I+U^{-1})^{-1}=(I+U)^{-1}+U(I+U)^{-1}=I$, so that $G_{i,j}(\tau)+G_{j,i}^*(\tau)=\delta_{i,j}$. With this knowledge, we can measure Chern band polarize correlation function $S\equiv\langle (\hat{N}_+ - \hat{N}_-)^2 \rangle/N^2$ for auxiliary field $l$ according to 
\begin{eqnarray}
	\langle (\hat{N}_+ - \hat{N}_-)^2 \rangle_l &=& \langle \sum_{\mathbf{k}_1,\tau_1,s_1} (c_{\mathbf{k}_1,+,\tau_1,s_1}^\dagger c_{\mathbf{k}_1,+,\tau_1,s_1} - c_{\mathbf{k}_1,-,\tau_1,s_1}^\dagger c_{\mathbf{k}_1,-,\tau_1,s_1}) \sum_{\mathbf{k}_2,\tau_2,s_2} (c_{\mathbf{k}_2,+,\tau_2,s_2}^\dagger c_{\mathbf{k}_2,+,\tau_2,s_2} - c_{\mathbf{k}_2,-,\tau_2,s_2}^\dagger c_{\mathbf{k}_2,-,\tau_2,s_2}) \rangle_l \nonumber\\
	&=& \sum_{\mathbf{k}_1,\tau_1,s_1} \sum_{\mathbf{k}_2,\tau_2,s_2} \langle c_{\mathbf{k}_1,+,\tau_1,s_1}^\dagger c_{\mathbf{k}_1,+,\tau_1,s_1} - c_{\mathbf{k}_1,-,\tau_1,s_1}^\dagger c_{\mathbf{k}_1,-,\tau_1,s_1} \rangle_l \langle c_{\mathbf{k}_2,+,\tau_2,s_2}^\dagger c_{\mathbf{k}_2,+,\tau_2,s_2} - c_{\mathbf{k}_2,-,\tau_2,s_2}^\dagger c_{\mathbf{k}_2,-,\tau_2,s_2} \rangle_l \nonumber\\
	&+& \langle c_{\mathbf{k}_1,+,\tau_1,s_1}^\dagger c_{\mathbf{k}_2,+,\tau_2,s_2} \rangle_l \langle c_{\mathbf{k}_1,+,\tau_1,s_1} c_{\mathbf{k}_2,+,\tau_2,s_2}^\dagger \rangle_l + \langle c_{\mathbf{k}_1,-,\tau_1,s_1}^\dagger c_{\mathbf{k}_2,-,\tau_2,s_2} \rangle_l \langle c_{\mathbf{k}_1,-,\tau_1,s_1} c_{\mathbf{k}_2,-,\tau_2,s_2}^\dagger \rangle_l \nonumber\\
	&=& \left(\sum_{\mathbf{k}_1,\tau} 2 \left[ G_{\mathbf{k}_1+,\mathbf{k}_1+}(\tau) - G_{\mathbf{k}_1-,\mathbf{k}_1-}(\tau)\right]^*  \right)^2 \nonumber\\
	&+& \sum_{\mathbf{k}_1,\mathbf{k}_2,\tau} 2 \left[ \left|  G_{\mathbf{k}_1 +,\mathbf{k}_2 +}(\tau) \right| ^2 +\left|  G_{\mathbf{k}_1 -,\mathbf{k}_2 -}(\tau) \right| ^2\right]
\end{eqnarray}
where we use $\pm$ in $c_{\mathbf{k}_1,\pm,\tau_1,s_1}^\dagger$ to label different Chern bands. One should be careful that the definition of Chern bands $\pm$ will exchange in Hamiltonian when involving the other valley.

\section{Zero ground state energy}
Zero ground state energy is important for sign bound analysis. We will introduce what is the sufficient conditions for zero ground state energy. At chiral limit in Chern basis, form factor $\lambda_{m,n,\tau}(\mathbf{k}, \mathbf{k+q+G})$ is diagonal about Chern bands and $\sum_{\mathbf{k}}\lambda_{+,+,\tau}(\mathbf{k}, \mathbf{k+G})=\sum_{\mathbf{k}}\lambda_{-,-,\tau'}(\mathbf{k}, \mathbf{k+G})$. If there is no attraction ($\gamma=0$), in Chern basis
\begin{eqnarray}
	H_{I}&=&\frac{1}{2 \Omega} \sum_{\mathbf{G}} \sum_{\mathbf{q} \in mBZ} V_1(\mathbf{q+G}) \delta \rho_{1,\mathbf{q+G}} \delta \rho_{1,\mathbf{-q-G}} \nonumber\\
	\delta \rho_{1,\mathbf{q+G}}&=&\sum_{\mathbf{k},m,\tau,s} \lambda_{m,m,\tau}(\mathbf{k}, \mathbf{k+q+G}) (c_{\mathbf{k}, m, \tau,s}^{\dagger} c_{\mathbf{k+q}, m, \tau,s}-\frac{\mu}{8}\delta_{\mathbf{q},0})
\end{eqnarray}
We define $\mu\equiv\nu+4$ for convenience here. For a certain integer filling tuned by $\mu$, we can find a ground state $\left| \psi_0 \right\rangle$ with whole Chern bands occupied by noticing a state satisfying $\delta\rho_{1,\mathbf{q+G}}\left| \psi_0 \right\rangle = 0$ for any $\mathbf{q+G}$ must be the ground state for this positive semi-definite Hamiltonian.
\begin{eqnarray}	
	\sum_{\mathbf{k},m',\tau',s'}\lambda_{m',m',\tau'}(\mathbf{k}, \mathbf{k+G}) - \mu \sum_{\mathbf{k}}\lambda_{m,m,\tau}(\mathbf{k}, \mathbf{k+G}) = 0
\end{eqnarray}
Here $m',\tau',s'$ mean filled Chern band, valley and spin indexes. One just requires the amount of filled bands is equal to $\mu$ to acquire zero energy ground state. This means \textit{all integer filling $\mu$ from 0 to 8 have zero energy ground state when $\gamma=0$}.

With finite attraction $\gamma>0$, we need one more satisfied restriction $\delta\rho_{2,\mathbf{q+G}}\left| \psi_0 \right\rangle = 0$ to obtain zero energy ground state. Explicitly, this requirement is
\begin{eqnarray}	
	\sum_{\mathbf{k},m',s'}\lambda_{m',m',\tau}(\mathbf{k}, \mathbf{k+G}) - \sum_{\mathbf{k},m'',s''}\lambda_{m'',m'',-\tau}(\mathbf{k}, \mathbf{k+G}) = 0
\end{eqnarray}
Here $m',s'$ mean filled Chern band, spin indexes in valley $\tau$ and $m'',s''$ are filled Chern band, spin indexes in valley $-\tau$. This restriction actually requires \textit{the same amount of filling bands in opposite valleys}, which rules out odd integer filling cases at $\gamma=0$.

For non-chiral $\gamma=0$ case, $\sum_{\mathbf{k}}\lambda_{+,+,\tau}(\mathbf{k}, \mathbf{k+G})=\sum_{\mathbf{k}}\lambda_{-,-,\tau'}(\mathbf{k}, \mathbf{k+G})$ cannot be promised. The zero energy ground state requirement becomes
\begin{eqnarray}	
	\sum_{\mathbf{k},m',\tau',s'}\lambda_{m',m',\tau'}(\mathbf{k}, \mathbf{k+G}) - \frac{\mu}{2} \sum_{\mathbf{k},m}\lambda_{m,m,\tau}(\mathbf{k}, \mathbf{k+G}) = 0
\end{eqnarray}
This means one needs \textit{full fill even opposite Chern bands}, which corresponds to even integer fillings. These zero ground state energy states will be the necessary condition for sign bounds analysis below.

\section{Ground state degeneracy and sign bounds}

Symmetry for no attraction ($\gamma=0$) chiral system is U(4)$\times$U(4) according to ~\cite{BAB3,BAB4}, where each U(4) corresponds to spin-valley freedom with the same Chern number bands sector. We use Young diagram for SU(4) to compute ground state degeneracy of different integer fillings. The ground state wave function irreducible representation for filling $\mu_+$ from 0 to 4 corresponding to 2 spins, 2 valleys within $+$ Chern bands sector can be expressed as Young diagram below respectively
\begin{center}
	$\mu_+=1, \ydiagram{4}\dots\dots\ydiagram{4}$ \\
	$\mu_+=2, \ydiagram{4,4}\dots\dots\ydiagram{4,4}$ \\
	$\mu_+=3, \ydiagram{4,4,4}\dots\dots\ydiagram{4,4,4}$ \\
	$\mu_+=4, \underbrace{\ydiagram{4,4,4,4}\dots\dots\ydiagram{4,4,4,4}}$ \\
	\qquad \quad N
\end{center}

The amount of normal Young tableau for SU(4) symmetry calculated by hook length formula~\cite{BAB4,hofmannFermionic2022} are
\begin{eqnarray}
	d_{\mu_+=0}(SU(4)) &=& 1 \nonumber\\
	d_{\mu_+=1}(SU(4)) &=& \prod_{j}\frac{4+j-1}{j} = \frac{\frac{(N+3)!}{3!}}{N!} = \frac{(N+3)(N+2)(N+1)}{3!} \nonumber\\
	d_{\mu_+=2}(SU(4)) &=& \prod_{i,j}\frac{4+j-i}{h_{i,j}} = \frac{\frac{(N+3)!(N+2)!}{3!2!}}{(N+1)!N!} = \frac{(N+3)(N+2)^2(N+1)}{3!2!} \nonumber\\
	d_{\mu_+=3}(SU(4)) &=& \frac{\frac{(N+3)!(N+2)!(N+1)!}{3!2!}}{\frac{(N+2)!(N+1)!N!}{2!}} = \frac{(N+3)(N+2)(N+1)}{3!} \nonumber\\
	d_{\mu_+=4}(SU(4)) &=& 1
\end{eqnarray}
By the way, the SU(2) Young diagram for $\mu_+=0, \mu_+=1, \mu_+=2$ have the same shape with the SU(4) above, but the amounts of normal Young tableau are different
\begin{eqnarray}
d_{\mu_+=0}(SU(2)) &=& 1 \nonumber\\
d_{\mu_+=1}(SU(2)) &=& \prod_{j}\frac{2+j-1}{j} = N+1 \nonumber\\
d_{\mu_+=2}(SU(2)) &=& 1
\end{eqnarray}

Now we are ready to count the total degeneracy for all integer filling $\mu$ from 0 to 8 (corresponding to $\nu$ from -4 to 4) for chiral no attraction ($\gamma=0$) TBG model by using SU(4) Young diagram.
\begin{eqnarray}
	d_{\mu=0} &=& d_{\{\mu_+=0,\mu_-=0\}}=1 \nonumber\\
	d_{\mu=1} &=& d_{\{\mu+=1,\mu_-=0\}}+d_{\{\mu_+=0,\mu_-=1\}}= \frac{(N+3)(N+2)(N+1)}{3} \nonumber\\
	d_{\mu=2} &=& 2d_{\{\mu_+=2,\mu_-=0\}}+d_{\{\mu_+=1,\mu_-=1\}}=\frac{(N+3)(N+2)^2(N+1)}{3!}+\frac{(N+3)^2(N+2)^2(N+1)^2}{3!3!} \nonumber\\
	d_{\mu=3} &=& 2d_{\{\mu_+=3,\mu_-=0\}}+2d_{\{\mu_+=2,\mu_-=1\}}=\frac{(N+3)(N+2)(N+1)}{3}+\frac{(N+3)^2(N+2)^3(N+1)^2}{3!3!} \nonumber\\
	d_{\mu=4} &=& 2d_{\{\mu_+=4,\mu_-=0\}}+2d_{\{\mu_+=3,\mu_-=1\}}+d_{\{\mu_+=2,\mu_-=2\}}=2+\frac{(N+3)^2(N+2)^2(N+1)^2}{3!3}+\frac{(N+3)^2(N+2)^4(N+1)^2}{3!3!2!2!} \nonumber\\
	d_{\mu=5} &=& 2d_{\{\mu_+=4,\mu_-=1\}}+2d_{\{\mu_+=3,\mu_-=2\}}=\frac{(N+3)(N+2)(N+1)}{3}+\frac{(N+3)^2(N+2)^3(N+1)^2}{3!3!} \nonumber\\
	d_{\mu=6} &=& 2d_{\{\mu_+=4,\mu_-=2\}}+d_{\{\mu_+=3,\mu_-=3\}}=\frac{(N+3)(N+2)^2(N+1)}{3!}+\frac{(N+3)^2(N+2)^2(N+1)^2}{3!3!} \nonumber\\
	d_{\mu=7} &=& 2d_{\{\mu_+=4,\mu_-=3\}}= \frac{(N+3)(N+2)(N+1)}{3} \nonumber\\
	d_{\mu=8} &=& d_{\{\mu_+=4,\mu_-=4\}}=1
\end{eqnarray}
When we consider $N\rightarrow\infty$, the degeneracy  with power of $N$ for different fillings can be seen as
$d_{\mu=0} \propto N^0,
d_{\mu=1} \propto N^3,
d_{\mu=2} \propto N^6,
d_{\mu=3} \propto N^7,
d_{\mu=4} \propto N^8,
d_{\mu=5} \propto N^7,
d_{\mu=6} \propto N^6,
d_{\mu=7} \propto N^3,
d_{\mu=8} \propto N^0$. Because the degeneracy for our reference system ($\mu=4$ case where there is no sign problem) is $d_{\mu=4} \propto N^8$, we can see the other integer filling have different polynomial sign bounds decay according to $\left\langle sign \right\rangle_b\equiv b_{\mu}=d_{\mu}/d_{\mu=4}$ as $b_{\mu=3,\mu=5} \propto N^{-1}, b_{\mu=2,\mu=6} \propto N^{-2}, b_{\mu=1,\mu=7} \propto N^{-5}, b_{\mu=0,\mu=8} \propto N^{-8}$. This is the first column result in our main text Tab. I.

One can verify the involving attraction case ($\gamma>0$) will break the U(4)$\times$U(4) to U(2)$\times$U(2)$\times$U(2)$\times$U(2), where original spin-valley U(4) is broken into two spin U(2). By introducing \textit{filling the same amount of bands in opposite valleys} restriction for zero energy ground states, we can write valley label $\tau$ explicitly for a given $\mu$ and use the SU(2) Young diagram results above to count the degeneracy of even integer fillings now
\begin{eqnarray}
d_{\mu=0} &=& d_{\{\mu_{\tau}=0,\mu_{-\tau}=0\}}=1 \nonumber\\
d_{\mu=2} &=& 4d_{\{\mu_{\tau}=1,\mu_{-\tau}=1\}}=4(N+1)^2 \nonumber\\
d_{\mu=4} &=& 4d_{\{\mu_{+,\tau}=2,\mu_{-,-\tau}=2\}}+4d_{\{\mu_{+,\tau}=2,\mu_{-,-\tau}=1,\mu_{+,-\tau}=1\}}+d_{\{\mu_{+,\tau}=1,\mu_{+,-\tau}=1,\mu_{-,\tau}=1,\mu_{-,-\tau}=1\}} \nonumber\\
&=& 4+4(N+1)^2+(N+1)^4 \nonumber\\
d_{\mu=6} &=& 4d_{\{\mu_{\tau}=3,\mu_{-\tau}=3\}}=4(N+1)^2 \nonumber\\
d_{\mu=8} &=& d_{\{\mu_{\tau}=4,\mu_{-\tau}=4\}}=1
\end{eqnarray}
At large $N$, sign bounds with attraction ($\gamma>0$) at chiral even integer filling are written as $b_{\mu=2,\mu=6} \propto N^{-2}, b_{\mu=0,\mu=8} \propto N^{-4}$, which is the result of the third column in our main text Tab. I.

It is well studied that non-chiral condition breaks U(4)$\times$U(4) into one U(4). We can count the ground state degeneracy of $\gamma=0$ case by SU(4) Young diagram
\begin{eqnarray}
d_{\mu=0} &=& 1 \nonumber\\
d_{\mu=2} &=& d_{\{\mu_{s,\tau}=2\}} = \frac{(2N+3)(2N+2)(2N+1)}{3!} \nonumber\\
d_{\mu=4} &=& d_{\{\mu_{s,\tau}=4\}} = \frac{(2N+3)(2N+2)^2(2N+1)}{3!2!} \nonumber\\
d_{\mu=6} &=& d_{\{\mu_{s,\tau}=6\}} = \frac{(2N+3)(2N+2)(2N+1)}{3!} \nonumber\\
d_{\mu=8} &=& 1
\end{eqnarray}
This gives the second column result of Tab. I in the main text, which are sign bounds $b_{\mu=2,\mu=6} \propto N^{-1}, b_{\mu=0,\mu=8} \propto N^{-4}$ for non-chiral $\gamma=0$ large $N$ case.

\section{Excitations from Exact solution}

To give an intuitive understanding for finite temperature physics, we compute the single particle, two kinds of charge-neutral excitonic exciations following the method in Ref.~\cite{BAB5}. Below, we first prove the single particle excitation states and charge neutral excitation states form closed orthogonal subspaces separately for the chiral no attraction ($\gamma=0$) Hamiltonian we use. Then, we derive the eigen excitation states by diagonalizing the Hamiltonian in the subspace.

We choose one full-filling Chern bands ground state as $\left| \psi_0\right\rangle $ where filled spin-valley bands labeled by $\sigma$ and empty bands labeled by $\sigma^\prime$. By defining raising operators and their Hermitian conjugate linking filled spin-valley bands and empty spin-valley bands in Chern number $\pm$ subspace
\begin{eqnarray}
	\Delta^\dagger_{\pm,\sigma^\prime,\sigma}=\sum_{\mathbf{k}}c_{\mathbf{k},\pm,\sigma^\prime}^\dagger c_{\mathbf{k},\pm,\sigma} \nonumber\\
	\Delta_{\pm,\sigma^\prime,\sigma}=\sum_{\mathbf{k}}c_{\mathbf{k},\pm,\sigma}^\dagger c_{\mathbf{k},\pm,\sigma^\prime}
\end{eqnarray}
we can check these operators commute with $\delta\rho_{\mathbf{q+G}}$ and also with the Hamiltonian. The commutation relations between these operators are
\begin{eqnarray}
	\left[ \Delta^\dagger_{\pm,\sigma_1^\prime,\sigma_1}, \Delta_{\pm,\sigma_2^\prime,\sigma_2}^\dagger\right] &=& \left[ \Delta_{\pm,\sigma_1^\prime,\sigma_1}, \Delta_{\pm,\sigma_2^\prime,\sigma_2}\right]=0 \nonumber\\
	\left[ \Delta^\dagger_{\pm,\sigma_1^\prime,\sigma_1}, \Delta_{\pm,\sigma_2^\prime,\sigma_2}\right]&=&\sum_{\mathbf{k}}c_{\mathbf{k},\pm,\sigma_1^\prime}^\dagger c_{\mathbf{k},\pm,\sigma_2^\prime} \delta_{\sigma_1,\sigma_2} - c_{\mathbf{k},\pm,\sigma_1}^\dagger c_{\mathbf{k},\pm,\sigma_2} \delta_{\sigma_1^\prime,\sigma_2^\prime}
\end{eqnarray}
For raising between empty bands or full bands, those operators will not change the state and only give a constant $\Delta^\dagger_{\pm,\sigma_1^\prime,\sigma_2^\prime} \left| \psi_0\right\rangle = const \left| \psi_0\right\rangle$ and $\Delta^\dagger_{\pm,\sigma_1,\sigma_2} \left| \psi_0\right\rangle = const \left| \psi_0\right\rangle$. We can then conclude all orthogonal ground states $\left| \psi_g\right\rangle$ can be expressed based on $\left| \psi_0\right\rangle$ and raising operators in a form below
\begin{eqnarray}
	\left| \psi_g\right\rangle = \prod_{\sigma^\prime,\sigma} (\Delta^\dagger_{\pm,\sigma^\prime,\sigma})^{m_{\pm,\sigma^\prime,\sigma}} \left| \psi_0\right\rangle
\end{eqnarray}
Here $m_{\pm,\sigma^\prime,\sigma}$ are non-negative integers. The reason why we can write down this form is basically commutation between $\Delta^\dagger_{\pm,\sigma_1^\prime,\sigma_2^\prime}$ and $\Delta^\dagger_{\pm,\sigma_3^\prime,\sigma_4}$ will be $\Delta^\dagger_{\pm,\sigma_1^\prime,\sigma_4}\delta_{\sigma_2^\prime,\sigma_3^\prime}$ which belongs to $\Delta^\dagger_{\pm,\sigma^\prime,\sigma}$ and the rest part after commutation will be $\Delta^\dagger_{\pm,\sigma_1^\prime,\sigma_2^\prime} \left| \psi_0\right\rangle = const \left| \psi_0\right\rangle$ which does not change the state and each $\Delta^\dagger_{\pm,\sigma^\prime,\sigma}$ will change the particle number while keeping momentum conservation in two bands so that different particle number distribution or different momentum conservation way states are orthogonal.

Now, we would like to check the orthogonal and close property for single particle excitation subspace $c^\dagger_{\mathbf{k_1},\pm,\sigma_1} \left| \psi_g\right\rangle$. Here $\sigma_1$ is not necessarily full-filled bands because bands can be partially filled in $\left| \psi_g\right\rangle$. Noticing $\left| \psi_0\right\rangle$ is single particle Kronecker product state, orthogonality can be easily checked by Wick's theorem. To be specific, only full contraction can give non-zero result $A\delta_{\mathbf{k_1},\mathbf{k_1^\prime}}\delta_{\sigma_1,\sigma_1^\prime}$ where $A$ is a constant. And close property can also be checked according to
\begin{eqnarray}
	&&H_I c^\dagger_{\mathbf{k_1},\pm,\sigma_1} \left| \psi_g\right\rangle = \left[ H_I, c^\dagger_{\mathbf{k_1},\pm,\sigma_1}\right] \left| \psi_g\right\rangle	\nonumber\\
	&=& \frac{1}{2 \Omega} \sum_{\mathbf{G}} \sum_{\mathbf{q} \in mBZ} V(\mathbf{q+G}) \delta \rho_{\mathbf{q+G}} \left[ \delta \rho_{\mathbf{-q-G}}, c^\dagger_{\mathbf{k_1},\pm,\sigma_1} \right] \left| \psi_g\right\rangle \nonumber\\
	&=& \frac{1}{2 \Omega} \sum_{\mathbf{G}} \sum_{\mathbf{q} \in mBZ} V(\mathbf{q+G}) \left[ \delta \rho_{\mathbf{q+G}}, \left[ \delta \rho_{\mathbf{-q-G}}, c^\dagger_{\mathbf{k_1},\pm,\sigma_1} \right] \right] \left| \psi_g\right\rangle \nonumber\\
	&=& \left[ \frac{1}{2 \Omega} \sum_{\mathbf{G}} \sum_{\mathbf{q} \in mBZ} V(\mathbf{q+G}) \lambda_{\pm,\sigma_1}(\mathbf{k_1}+\mathbf{q}+\mathbf{G},\mathbf{k_1}) \lambda_{\pm,\sigma_1}(\mathbf{k_1},\mathbf{k_1}+\mathbf{q}+\mathbf{G}) \right]  c^\dagger_{\mathbf{k_1},\pm,\sigma_1} \left| \psi_g\right\rangle
\end{eqnarray}
This means $H_I$ only projects such a single particle excitation state to linear combination of single particle excitation states. Then one can diagonalize $H_I$ in this subspace to derive eigen single particle excitation. It is interesting to notice that the single particle excitation has nothing to do with the constant $\mu$ because this constant does not contribute to the commutation relation. Then the single particle excitation here will be exactly same for any integer chemical potential where it has zero energy and the single hole excitation is exactly the same with particle due to particle hole symmetry. This is consistent with STM experiment where LDOS slowly change with $\nu$~\cite{kerelskyMaximized2019}

For charge neutral two particle excitation $c^\dagger_{\mathbf{k_1},\pm,\sigma_1} c_{\mathbf{k_2},\pm,\sigma_2} \left| \psi_g\right\rangle$ or $c^\dagger_{\mathbf{k_1},\pm,\sigma_1} c_{\mathbf{k_2},\mp,\sigma_2} \left| \psi_g\right\rangle$, one need to exclude self-contraction within one side which gives $\delta_{\mathbf{k_1},\mathbf{k_2}}\delta_{\mathbf{k_1^\prime},\mathbf{k_2^\prime}}$ but not $\delta_{\mathbf{k_1},\mathbf{k_1^\prime}}\delta_{\mathbf{k_2},\mathbf{k_2^\prime}}$ as we want. For $c^\dagger_{\mathbf{k_1},\pm,\sigma_1} c_{\mathbf{k_2},\mp,\sigma_2} \left| \psi_g\right\rangle$, this can be excluded automatically because $\pm\neq\mp$ will force contraction at the same side to be zero. For $c^\dagger_{\mathbf{k_1},\pm,\sigma_1} c_{\mathbf{k_2},\pm,\sigma_2} \left| \psi_g\right\rangle$, only when $\mathbf{k_1}=\mathbf{k_2}$, $\sigma_1=\sigma_2$ contraction at the same side can happen. If we ignore this one point then the left states again form an orthogonal and close subspace

\begin{eqnarray}
	&&H_I c^\dagger_{\mathbf{k},\pm,\sigma_1} c_{\mathbf{k+p},\pm,\sigma_2} \left| \psi_g\right\rangle = \left[ H_I, c^\dagger_{\mathbf{k},\pm,\sigma_1} c_{\mathbf{k+p},\pm,\sigma_2} \right] \left| \psi_g\right\rangle	\nonumber\\
	&=& \frac{1}{2 \Omega} \sum_{\mathbf{G}} \sum_{\mathbf{q} \in mBZ} V(\mathbf{q+G}) \left[ \delta \rho_{\mathbf{q+G}}, \left[ \delta \rho_{\mathbf{-q-G}}, c^\dagger_{\mathbf{k},\pm,\sigma_1} c_{\mathbf{k+p},\pm,\sigma_2} \right] \right] \left| \psi_g\right\rangle \nonumber\\
	&=& \frac{1}{2 \Omega} \sum_{\mathbf{G}} \sum_{\mathbf{q} \in mBZ} V(\mathbf{q+G}) [ \lambda_{\pm,\sigma_1}(\mathbf{k},\mathbf{k}+\mathbf{q}+\mathbf{G})\lambda_{\pm,\sigma_1}(\mathbf{k}+\mathbf{q}+\mathbf{G},\mathbf{k}) c^{\dagger}_{\mathbf{k},\pm,\sigma_1} c_{\mathbf{k}+\mathbf{p},\pm,\sigma_2} \nonumber\\
	&&- 2\lambda_{\pm,\sigma_1}(\mathbf{k}+\mathbf{p},\mathbf{k}+\mathbf{p}+\mathbf{q}+\mathbf{G})\lambda_{\pm,\sigma_2}(\mathbf{k}+\mathbf{q}+\mathbf{G},\mathbf{k})c^{\dagger}_{\mathbf{k}+\mathbf{q},\pm,\sigma_1} c_{\mathbf{k}+\mathbf{q}+\mathbf{p},\pm,\sigma_2} \nonumber\\
	&&+ \lambda_{\pm,\sigma_2}(\mathbf{k}+\mathbf{p}+\mathbf{q}+\mathbf{G},\mathbf{k}+\mathbf{p})\lambda_{\pm,\sigma_2}(\mathbf{k}+\mathbf{p},\mathbf{k}+\mathbf{p}+\mathbf{q}+\mathbf{G})  c^{\dagger}_{\mathbf{k},\pm,\sigma_1} c_{\mathbf{k}+\mathbf{p},\pm,\sigma_2} ] \left| \psi_g\right\rangle
\end{eqnarray}

\section{Supplemental figures}

We list supplemental figures here. The converging behavior of average sign with temperature at chiral $\gamma=0,\nu=1$ case is shown in Fig.~\ref{fig:fig4}(a). One can see at simulation temperature $T=1$ meV which the FIG. 1 in the main text used, the sign converge to the finite value both for $L=3, L=4$. In contrast, for non-chiral $\gamma=0,\nu=1$ case as shown in Fig.~\ref{fig:fig4}(b), the average sign shows the usual exponential decay at low temperature. It is also worth to notice that the topological phase transition temperature corresponds to a local curvature maximum of the average sign, which should be the case because the second-order derivative of sign bounds corresponds to second-order derivative of partition function where there should be a divergence for second-order phase transition. For finite size system, this divergence is captured by a local curvature maximum. Besides, above the phase transition temperature at non-chiral case as shown in Fig.~\ref{fig:fig4}(b), we can see there is also an anomalous lift of average sign as the lift at chiral. We speculate at this temperature, the fluctuation between different Chern bands is allowed so that the distinction between chiral and non-chiral is fuzzy. The entropy coming from huge degeneracy of chiral case contributes to the lift of the partition function as well as the average sign.
\begin{figure}[htp!]
	\includegraphics[width=0.8\columnwidth]{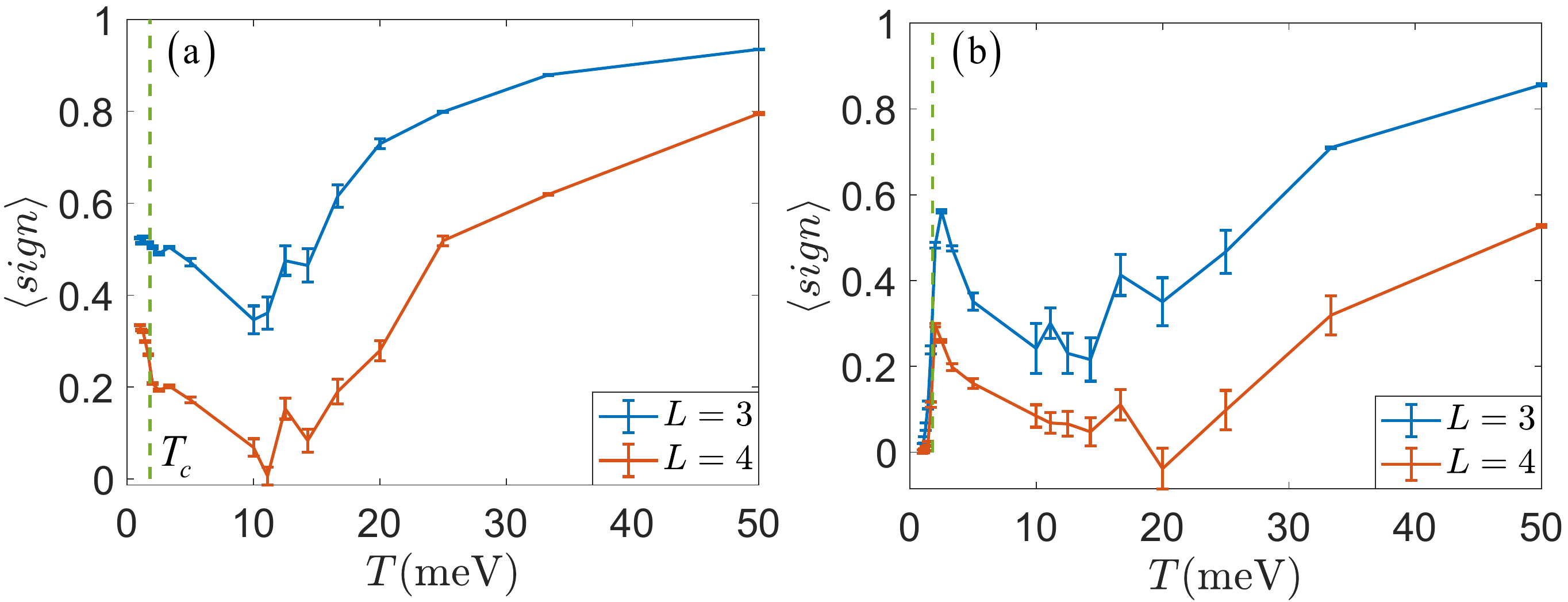}
	\caption{(a) Average sign versus temperature for chiral $\gamma=0,\nu=1$ case. Dash line indicates topological phase transition temperature $T_c\sim1.8$ meV (b) Average sign versus temperature for non-chiral $\gamma=0,\nu=1$ case.}
	\label{fig:fig4}
\end{figure}

We also measure the Chern band polarization correlation function $S\equiv\langle (\hat{N}_+ - \hat{N}_-)^2 \rangle/N^2$ at chiral $\gamma=0,\nu=2$ to confirm the Chern number keeps zero here as shown in Fig.~\ref{fig:fig5}(b), in contrast with the chiral $\gamma=0,\nu=1$ case in Fig.~\ref{fig:fig5}(a) which is the case discussed in the main text.
\begin{figure}[htp!]
	\includegraphics[width=0.8\columnwidth]{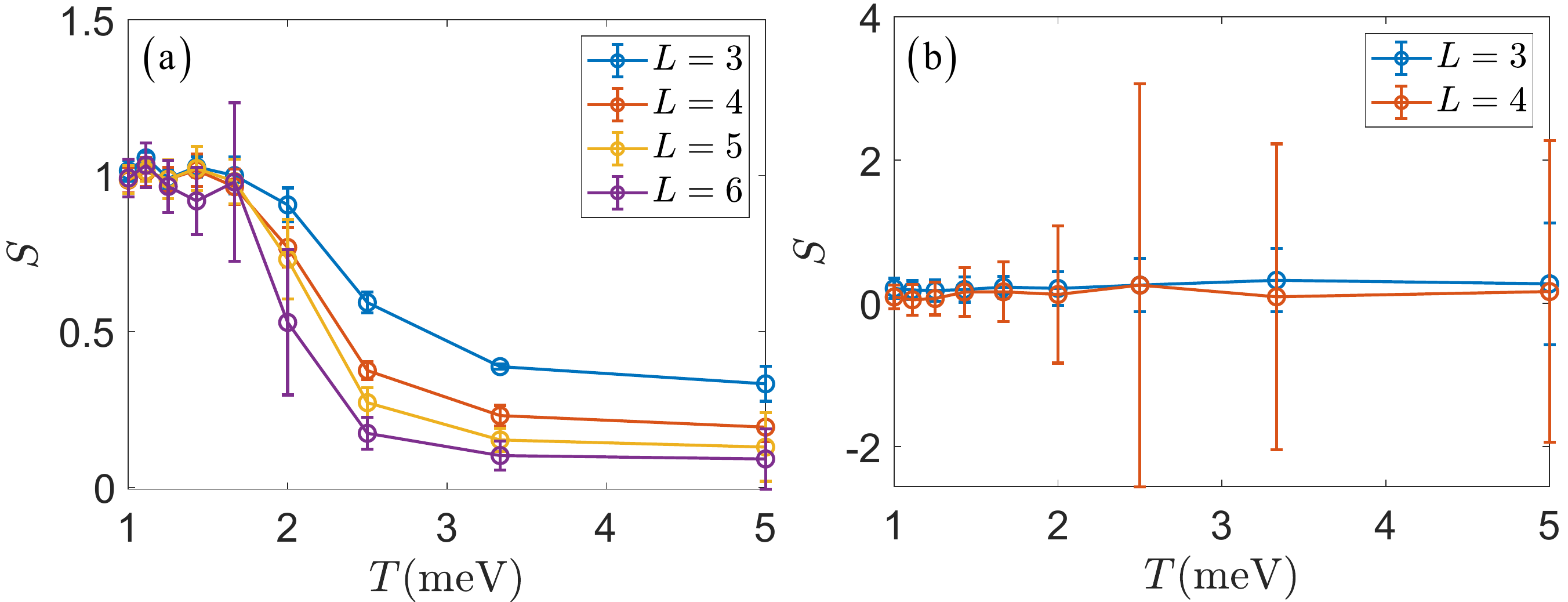}
	\caption{(a) Chern band polarization correlation function versus temperature for chiral $\gamma=0,\nu=1$ case. (b) Chern band polarization correlation function versus temperature for chiral $\gamma=0,\nu=2$ case.}
	\label{fig:fig5}
\end{figure}

\end{widetext}

\end{document}